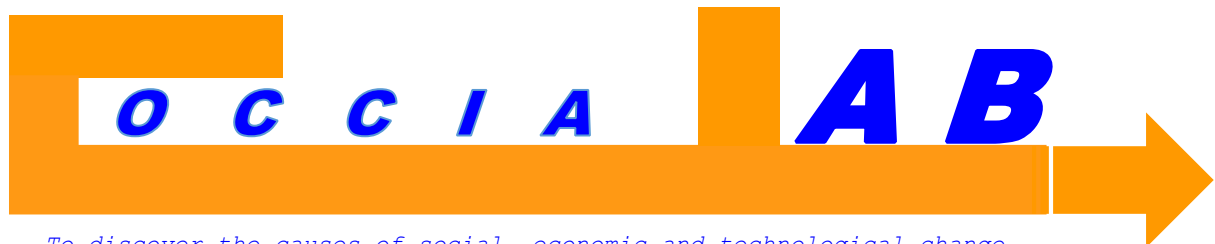

*To discover the causes of social, economic and technological change*

CocciaLAB Working Paper 2018 – No. 30

# Critical analysis of human progress: Its negative and positive sides in the late-capitalism

Mario COCCIA & Matteo BELLITTO

CNR -- NATIONAL RESEARCH COUNCIL OF ITALY

&

ARIZONA STATE UNIVERSITY

# Critical analysis of human progress: Its negative and positive sides in the late-capitalism

*Mario Coccia* [1,2] *& Matteo Bellitto*[1]
1. CNR -- National Research Council of Italy
2. Arizona State University

*E*-mail: mario.coccia@cnr.it, matteo.bellitto@gmail.com

Current Address: CocciaLAB at CNR -- National Research Council of Italy
Via Real Collegio, 30, 10024-Moncalieri (Torino), Italy

Mario Coccia ORCID: http://orcid.org/0000-0003-1957-6731

**Abstract.** The concept of progress has characterized human society from millennia. However, this concept is elusive and too often given for certain. The goal of this paper is to suggest a general definition of human progress that satisfies, whenever possible the conditions of independence, generality, epistemological applicability and empirical correctness. This study proposes, within a pragmatic approach, human progress as an inexhaustible process driven by an ideal of maximum wellbeing of purposeful people which, on attainment of any of its goals or objectives for increasing wellbeing, then seek another consequential goal and objective, endlessly, which more closely approximates its ideal fixed in new socioeconomic contexts over time and space. The human progress, in the global, capitalistic, and post-humanistic Era, improves the fundamental life-interests represented by health, wealth, expansion of knowledge, technology and freedom directed to increase wellbeing throughout the society. These factors support the acquisition by humanity of better and more complex forms of life. However, this study shows the inconsistency of the equation economic growth= progress because human progress also generates - during its continuous process without limit– negative effects for human being, environment and society.

**Suggested citation:**
Coccia M., Bellitto M. 2018. Critical analysis of human progress: Its negative and positive sides in the late-capitalism, *Working Paper CocciaLab* n. 30, CNR -- National Research Council of Italy, Torino (Italy)

Acknowledgement. *Mario Coccia is grateful to financial support from the CNR - National Research Council of Italy for his visiting at Arizona State University (Research Grant n. 0072373-2014 and n. 0003005-2016) where this research started in 2016. The authors thank Prof. Enrico Filippi for fruitful suggestions and comments on a preliminary draft of this paper. Authors declare that they have no relevant or material financial interests that relate to the research discussed in this paper.*

## Contents

**Purpose of this study**

This paper has two goals. The first is to define human progress. The second is a critique of the universal idea of "economic growth = progress" because the concept is stratified in manifold factors and includes both positive and negative dimensions in society.

The crux of the study here is rooted in the concept of progress in social sciences and a brief background is useful to understand and clarify it. The origin of the concept is the Latin *progressus,* derivation from *progrĕdi* "to walk forth, to advance". Progress is a process towards new and different phases that should be better. For this reason, the concept of progress has also been associated with the notion of evolution, though the terms are not synonym and cannot be used interchangeably (Woods, 1907, p. 780). Human progress is driven by science advances, technological change, efficient use of energy production, democratization, etc. In fact, scientific discoveries from Seventeenth and Eighteenth centuries by Galileo (1564-1642), Kepler (1571-1630) and Newton (1642-1727), together with the French Revolution (1789), were the flywheel for the second Industrial Revolution (1856-1878) and creation of new nations that generated a general growth of employment and production in Western-style economies (Coccia, 2005a; 2007). New scientific achievements and subsequent technological innovations pushed economy and redesigned the socioeconomic structure of countries (Nisbet, 1994, Usher, 1954; cf., Coccia, 2010, 2012a, 2014a, 2014b, 2016, 2016a; Coccia and Wang, 2015, 2016)[1]. Europe and the Western world discovered themselves as an economic and industrial engine, driven by their middle class which was born at that time and gained a new political weight in society (Coccia, 2017, 2018; Ruttan, 2001; Singer, 1956; Rae, 1834). The economic boost of Industrial Revolution in Europe and North America was adopted as the main indicator of progress and the concept of human progress has started to dawn as an *ex post* justification (Nisbet, 1994; Seligman, 1902).

---

[1] Cf. also Coccia and Finardi, 2012; Coccia et al., 2010; Coccia, 2002; 2005c; Coccia, 2012d, 2014f; 2015c; 2018a; Coccia and Wang, 2015; Calabrese et al., 2005, 2002; Calcatelli et al., 2003.

Hence, after these main facts of economic history, the meaning and perception of progress has been linked, more and more, to new science and technology's economic effects rather than social criteria. This is an evidence of the strong connection between doctrines of progress (such as Evolutionism, Positivism and Historical materialism) and historical events (Woods, 1907; De Greef, 1895).

Woods (1907, p. 815) argues that:

> a valid conception of progress must, first of all, depend upon results drawn wholly from an inspection of reality. In the second place, it must present not merely a descriptive or genetic account of the course of human evolution through successive eras, but a distinctly evaluative -that is, a teleological- formulation of the worthful elements in this evolution. And finally, in the endeavor to frame such a criterion, one must be content with nothing less than an impartial and comprehensive survey of the whole of human life.

In order to suggest a comprehensive definition of human progress suitable for clarifying the general development of societies over time and space, next section presents a theoretical framework based theorists in the Nineteenth and Twentieth Century.

**Human progress in philosophy: a theoretical framework**

Western culture has always dealt with the idea of progress, from Greek mythology — e.g. the myth of Prometheus — to the Contemporary Age (Small, 1905; Flint 1874). The idea of human progress was strongly promoted by Enlightenment and its thinkers, who claimed that, through the power of reason, people can upgrade and improve their knowledge in order to master the environment and save themselves from ignorance and poverty (cf., Wagner, 2015; West Churchman and Ackoff, 1950). Thus philosophical advocates of progress assume that the human condition has improved over the course of history and will continue to improve (Flint, 1884). Doctrines of progress appeared in the Eighteenth century in Europe embodied the optimism of that period. Subsequently, faith in human progress developed in the Nineteenth century is due to philosophers - like Comte and Positivism, Spencer - and Utilitarianism (Nisbet, 1994).

Comte (1875), the father of Positivism, is the first to use the term "Sociology" to describe the scientific treatment of human societies and their development. Thus he gave sociology its content in addition to its name. In Comte's thought, intellectual improvement drives progress and it should be understood as change in the form of explanation employed by individuals looking for understanding the world. Comte (1875) wanted to work the problem out systematically. In his system, rebuilding the development of society means to propose a real philosophy of history marked in three moments that reflect the law of the three stages: the theological, the metaphysical, and the positive one (Comte, 1875). In the theological stage, scientific explanation is driven by the assumption that natural events are caused by divinities; in the second phase, the theoretical one, phenomena are explained by referring to the abstract essences that entities are supposed to possess; in the last and so-called positive stage, scientific laws allow to explain individual phenomena and to master the environment. In this theoretical stream, the first principle of the Positivism is the following: rejecting the search for the ultimate reason of things to consider the facts and their actual laws (Comte, 1896). The recourse to facts, to experimentation, to the proof of reality, is what allows us to get out of speculative discourses and the search for the absolute, accepting the limits inherent to reason and therefore the relativity of knowledge (West Churchman and Ackoff, 1950).

Subsequently, the concept of human progress started to fail together with the blind faith in rationality (Woods, 1907). Moreover, the criticism of the Divine Plan is lacking in applicability (Small, 1905; cf., Flint, 1874). A criticism is also against the conception of natural law as criterion of social and individual progress.

The concept of progress, during the latter half of the 19$^{th}$ century, has been affected by theories of evolution (Woods, 1907). In this context, the main purpose of Spencer's theory is the construction of a huge philosophical system which, starting from biology — following Darwin's works, but also Lamarck's ones — extends and includes psychology, sociology, ethical and educational theories. Unlike Darwin (1859), which reduces evolutionism to the biological sphere, Spencer (1851, 1857) supports the idea of a "cosmic evolutionism". In short, there are different evolutions: in addition to the organic one, there is an inorganic evolution, and a subsequent super-organic one. The latter refers to the man and his realizations: culture, institutions, and society (Woods, 1907). This approach can support a theory of human progress based on the assumption that man, as part of nature, follows the same evolutionary process: "Progress, therefore, is not an accident, but a necessity. Instead of civilization being artificial, it is part of nature; embryo or the unfolding of a flower. The modifications mankind have undergone, and are still undergoing and provided the human race, and the constitution of things remains the same, those modifications must end in completeness" (Spencer, 1851, Pt. I, Ch. 2). Thus it emerges clearly the fundamental approach of his thought, which means evolution and progress as the universal laws of life and cosmos, according to a general and progressive movement that runs from the homogeneous to the heterogeneous. In fact, Spencer (1857) pointed out a process, from an indefinite homogeneity to a definite coherent heterogeneity, which is associated with a complete integration of the whole and by increased interdependence of the parts (Woods, 1907, p. 795).

In particular, Spencer (1857, pp. 446-447) argues:

> It is settled beyond dispute that organic progress consists in a change from the homogeneous to the heterogeneous … Now, we propose in the first place to show that this law of organic progress is the law of all progress. Whether it be in the development of the Earth, in the development of Life upon its surface, in the development of Society, of Government, of Manufactures, of Commerce, of Language, Literature, Science, Art, this same evolution of the simple into the complex, through a process of continuous differentiation, holds throughout. From the earliest traceable cosmical changes down to the latest results of civilization, we shall find that the transformation of the homogeneous into the heterogeneous is that in which Progress essentially consists.

Subsequently, Spencer (1902) changes his approach from the "law of progress" to the "law of evolution". Spencer (1902, p. 253) stated that:

> There is another form under which civilization can be generalized. We may consider it as a progress towards that constitution of man and society required for the complete manifestation of every one's individuality. … the full happiness of each, and therefore to the greatest happiness of all. Hence, in virtue of the law of adaptation, our advance must be towards a state in which this entire satisfaction of every desire, or perfect fulfilment of individual life, becomes possible.

So, Spencer (1902) has stressed the realization of individual happiness through an age-long process of adaptation. In this context, the idea of social progress is also specified by De Greef (1895, pp. 337-255):

> Progress implies a perfecting of the social organization, a perfecting such that the new society represents a variety superior to the mother society. This superiority should appear in a greater structure, and one, moreover, that is more differentiated and better coordinated, and in a corresponding vital functioning.…Social progress is directly proportional to the mass, to the differentiation, and to the co-ordination of the social elements and organs…. Neither the development nor the amount of wealth, of population, of art of knowledge, constitutes in itself progress, but only the conditions which may favor it; organization and progress are synonymous; they are substitutes the one for the other, as money is for merchandise.

Woods (1907, p. 797), analyzing these scholars, argued that: "Human progress is thus regarded as the necessary outcome of a universal biological process conceived, if only grasped with sufficient comprehensiveness, as working out the noblest results in every branch of human activity".

Gumplowicz (1883, p. 193 *et passim*) claimed the evolutionary incident of struggle among themselves: "the heterogeneous ethical and even social groups and communities carry forward the movement of history". In this respect Gumplowicz (1883) has a similar position to Nietzsche (1874) that also uses the primarily biological concept of struggle. However, terms as struggle, conflict, survival, and adaptation are important but cannot explain the true nature of human progress. "Progress is essentially a teleological idea, an idea of value. It cannot, therefore, be reduced to a formulation in terms of mechanism" (Woods, 1907, p. 800). Nietzsche (1874) critically addresses the concept of history and the conception of its linear development within European culture. He emphasizes precisely how different attitudes exist in history: his targets are on the one hand historicism, which rests on the idea that man is the result of the history that precedes him, and on the other hand the theoretical attitude that implies the possibility of an objective knowledge of man. The main aspect of man, underlines Nietzsche (1874), nevertheless resides in his subjectivity. Hence, a clear change of perspective from the philosophies of history promoted by Positivism and Utilitarianism is thus apparent (Nietzsche, 1874, 1954). The Twentieth century opens with a criticism to the doctrine of progress containing denials of the claim that the human condition is improving.

A very strong criticism of the idea of progress, based on the intrinsically negative aspects of scientific and technological progress, comes from the Frankfurt School, which distances itself from classical Marxism. According to this school, the domain brought by science is actually a form of slavery[2]. Horkheimer argues that: "the world is about to get rid of morality, becoming total organization that is total destruction. *Progress* tends to culminate in a catastrophe" (Horkheimer and Adorno, 1947, p. 11, Italics added). Adorno (1951, *passim*) denounces the "bad conscience" of progress, which while free destroys, also distrusts the extreme anti-progressism, which can overturn in irrationalism.

In *Minima Moralia*, he mentions that the writings by Benjamin over 1892–1940 period are an inspiration to him. In fact, Benjamin (1969) in the *Theses on History* offers a criticism of the Hegelian and Marxian philosophy of history[3]. The ninth thesis speaks for itself (Benjamin, 1969, pp. 257–8):

> A Klee painting named 'Angelus Novus' shows an angel looking as though he is about to move away from something he is fixedly contemplating. His eyes are staring, his mouth is open, his wings are spread. This is how one pictures the angel of history. His face is turned toward the past. Where we perceive a chain of events, he sees one single catastrophe which keeps piling wreckage upon wreckage and hurls it in front of his feet. The angel would like to stay, awaken the dead, and make whole what has been smashed. But a storm is blowing from Paradise; it has caught in his wings with such violence that the angel can no longer close them. This storm irresistibly propels him into the future to which his back is turned, while the pile of debris before him grows skyward. This storm is what we call progress.

---

[2] cf. also technological dependency in the studies by McLuhan, 1988; Ellul, 1964; Marcuse, 1977.
3 For Hegel's theories see Hegel (1902, 1807, 1837).

This philosophical and historical excursus about the concept of progress shows that it is elusive and includes manifold dimensions in society that are briefly discussed in the next section.

**Manifold dimensions of human progress**

Human progress can be considered as a system of manifold forces directed to improve wellbeing in society. Some of the most important driving forces of human progress, without pretending to be comprehensive are: science, technology, economic growth, energy, and democratization.

*Science*

What is meant by progress in science? During the Twentieth century its role in society has grown so much that it has become functional to civil and military state institutions, as well as a central position in world production, technological and economic processes (cf., Ruttan, 2001, 2006; Coccia, 2005, 2015; 2017b, 2017g, 2018; Coccia and Wang, 2016; Stephan, 1996). However, the traditional cumulative view of scientific knowledge was effectively challenged by many philosophers of science over 1950s-1970s period (Popper, 1959; cf., Kuhn, 1970; Lakatos, 1978). In addition to the question of progress in science, another problem is represented by the role that science itself plays within contemporary society (Coccia, 2012b). For instance, Lyotard (1979) argues that the state is willing to spend a lot of money in R&D investments to make science a driver of economic growth in society: this allows it to acquire credibility and to create public consensus that serve its decision-making bodies (cf., Coccia,

2010b, 2012b, 2013, 2017a, 2017c; 2017i; 2017l; 2017m; 2017n)[4]. Science is indeed a multi-layered system involving a community of scientists engaged in international research collaboration using methods of inquiry in order to produce new knowledge and/or science advances within and between scientific fields (Coccia and Wang, 2016). Thus the notion of science may refer to different shapes: to a social institution, to research process, to method of inquiry, and to scientific knowledge (Coccia, 2006, 2014, 2014c, 2014d, 2015a, 2017h). During the late capitalism, science is considered a driver of technology 'science-push model' and R&D investments by governments and enterprises produce technology and, as a consequence, economic growth and social change (Coccia, 2017h).

*Technology*

Technology is another main dimension of human progress and is strictly linked to science that thus becomes the precondition for technological development (Coccia, 2010a, 2012b, 2017e, 2017f; Basalla, 1988). Through technology — both in its anthropological meaning of "human activity and means to an end" and in the Heideggerian terminology of "opening" — people fill the gap from and manipulate the environment and in general, the nature (Heidegger, 1954; Coccia, 2015a). Current scientific research in artificial intelligence and computational approaches to problems, it can support new processes of scientific discovery and technology for human progress (Thagard, 1988). From this aspect we can see how the aims of science and technology refer to the original question linked to progress, that is the will of Western society to free itself from obscurantism and together from material poverty (Wagner, 2015; Coccia, 2013). In short, science and technology (S&T) have been the driving forces of societies over the course of time. It is therefore clear that S&T cannot be isolated from the economic context in which they take place and from the effects they produce on socioeconomic systems (Coccia, 2014d, 2015).

---

[4] Cf., Benati and Coccia, 2017; Coccia, 2001, 2003, 2004, 2008, 2014e; Coccia and Cadario, 2014; Coccia et al., 2015; Coccia and Rolfo, 2009, 2010, 2013; Cavallo et al., 2014, 2015; Ferrari et al., 2013.

The current growth of scientific-technological system has assumed rapid speed in the history of humanity and has induced profound social changes worldwide (Mesthene, 1971, Coccia, 2016).

*Economic growth*

Technological change and general purpose technologies support a third main element of the progress: the economic growth (Lucas, 1988; Romer, 1994; Coccia, 2007, 2018, 2017, 2015b, p. 61ff). Modern literature considers the term "human development" as an expansion of human capabilities, a widening of choices, 'an enhancement of freedom, and a fulfilment of human rights' (Srinivasan, 1994; Anand and Sen, 2000; Welzel et al., 2003). The economic view of the progress is the base for the Human Development Index (HDI) that is designed as a composite statistic index of life expectancy, education, and per capita income indicators, used to rank countries into four tiers of human development (Streeten, 1994; cf. Human Development Report, 2013). In general, countries with high HDI have a high level of economic growth, such as Norway, Australia, Canada, Switzerland, etc., whereas countries with low HDI have a low economic growth, such as Afghanistan, Iraq, Liberia, etc. (Norris, 2015).

*Energy*

Another main element of progress is energy forces and efficient energy system (Coccia, 2010c, p. 1330). The second Industrial Revolution has spread the electricity generating a need of natural resources to support energy supply and use in the late capitalism, more and more focused on mass consumption that is generating a consistent social, economic and environmental change. In fact, the huge energy production, associated with industrialization and consumption, has driven both human development and environmental effects, such that several scholars debate the concept of a new geological epoch called "Anthropocene" (Crutzen and Stoermer, 2000; Zalasiewicz et al., 2011; cf., Coccia, 2015b). But progress, based on energy supply, also goes through a conscious development of its environmental resources and the development of policies to protect them. Increasingly, in the Twentieth century it relied unconditionally on non-renewable energy sources, in order to support the sustainable production of consumer goods. A rethinking of energy use is generating the development of a critical conscience about the theme of sustainability in the Western world. Hence, both advanced and emerging economies have the hard task of commensurate their economic growth to its sustainability in human and environmental terms.

*Democratization*

Improvements in science, technology and energy in society are both the cause and consequence of the economic development in which they are expressed (Coccia, 2010, 2012b, 2014c, 2014d). But should we infer that a more prosperous, capital- and technology-based economy leads to an improvement in the political system and, more generally, its progressive democratization? It is important to try to answer this question keeping in mind that a large part of historians think that never before in the Nineteenth century it was possible to witness a general implementation of science, technology, and economic growth associated with a progressive democratization in society that supports the social background of human progress (cf., Lipset Seymour, 1959). In fact, Western society has been able to take advantage of these developments and, in parallel, has developed a greater tendency towards democratization associated with higher innovative outputs (Coccia, 2010). Modelski and Perry III (2002, p. 370-72) show that democracy is a:

> a process that had taken 120 years to travel from the position of 10% saturation (about 1880) to reach the flex-point of 50% about the year 2000 could also be expected to take a similar length of time to reach the 90% level (the estimated time constant of that process being 228 years, …). The earlier rate of increase in the size of the democratic community is likely to become more difficult to achieve and will decline over time…. Democratization represents possibly the most important developmental trend in the world today, and it bears continuous watching but it does not justify excessive optimism.

In principle, with due caution, it can be said that the economically healthier societies, with higher innovative outputs, are also the most democratic (cf., Acemoglu et al., 2008; Coccia, 2010; Modelski and Perry III, 1991, 2002). In fact, Coccia (2010, p. 248, original emphasis) shows that: "most free countries, measured with liberal, participatory, and constitutional democracy indices, have a higher level of technology than less free and more autocratic countries. … 'democracy richness' generates a higher rate of technological innovation with fruitful effects for the wellbeing and wealth of nations. These findings and predictions lead to the conclusion that policy makers need to be cognizant of positive associations between democratization and technological innovation paths in order to support the modern economic growth and future technological progress of countries".

Overall, then, progress includes a complex set of dimensions and there is a difference between the modern conception of progress oriented to an unconditional growth and the contemporary concept, oriented towards its critical re-discussion considering the sustainability.

**A new definition and critique of human progress**

Considering the arguments above, human progress can be defined as an *inexhaustible process driven by certain ideal objectives of wellbeing and satisfaction to be achieved*– presumably achievable in the short term – which on attainment of any of goals or objectives, the purposeful people seek other consequential objectives, endlessly, for improving general wellbeing towards the ideal of maximum satisfaction, in a sustainable society over time and space.

In short, human progress is an inexhaustible process driven by an ideal of maximum wellbeing of people which, on attainment of any of its goals or objectives, then seeks another consequential goal and objective endlessly for improving wellbeing and satisfaction adapted in sustainable society. The human progress has a concept of perfection and pursues it systematically over time and space; that is, in interrelated steps. The horizon of human progress is a regulatory ideal for improving the human condition. Progress is therefore a tension and is in fact unattainable. Liberty is one of the conditions of this infinite deployment of means.

Hence, the ideal of human progress as time approaches infinity is the expansion of scientific knowledge, technology, energy production and culture directed to the improvement of health, communications and transport technology, wealth, sociability and freedom in society. These factors support the acquisition by humanity of better and more complex forms of life. In general, the engine of the human progress seems to be an invisible hand that guides human nature towards self-determination for improving wellbeing and generating social benefits widespread in society over the long run (cf., Smith, 1761). Spencer (1902, p. 253) argued that: "the full happiness of each, and therefore to the greatest happiness of all".

In particular, the idea of progress is based on progressive satisfaction of human wants in all their ramifications and complexities. It is this inner kernel of human satisfactions which gives character to the whole account of social evolution; which is interpreted, not in terms of mechanism, … but of purpose (Woods, 1907, p. 816). Some scholars argue that human progress is driven by: "harmonious satisfaction of universally appreciated and highly developed interests diffused throughout the society" (Woods, 1907, p. 816). The fundamental life-interests in society, as said, are health, wealth, sociability, knowledge, beauty, and rightness interests, etc. (cf., Small, 1905, p. 682).

This definition of human progress takes place within a pragmatist horizon, where thought is always oriented towards goals, endlessly, and an active intervention on reality. Darwin and his theory of evolution have influenced Utilitarianism and Positivism concept of progress. But the American pragmatism has also inherited from the Darwinian conception, especially in Dewey's thought that the human being is always in a dynamic and conflictual relationship with the environment (Dewey, 2008, *passim*). This is why the new definition here echoes its formulations and why human progress can be seen as an ideal-seeking system directed to improve wellbeing in sustainable society (cf., Ackoff, 1971). In this context, the idea of human progress suggested here has an inner teleological foundation based on final cause and purposefulness[5] that are concepts necessary for understanding certain natural behavior of human society. Teleological behavior of human progress here seems to be also driven by a collective behavior controlled by negative feedback (cf., Rosenblueth et al. 1943, pp. 23-24).

All in all, if we were then to identify the fundamental directions of progress that circumscribe the field of activities by which man is able to overcome himself and to realize himself as such, we would find them precisely – and as previously anticipated – in science and technology, that are based on societies with higher democratization and sustainable economic growth. These basic indicators can be used to assess some vital characteristics of human progress in society. However, the concept of human progress also includes negative dimensions that will be discussed in the next section.

**Positive and negative sides of human progress in society: a critique**

From the point of view of liberty – foundation and presupposition of every democracy – classic dualism is proposed: positive liberty versus negative liberty. Positive liberty is involved in the answer to the questions: "What, or who, is the source of control or interference that can determine someone to do, or be, this rather than that?" (Berlin, 1958, pp. 15-16). The two questions are clearly different, even though the answers to them may overlap. Instead, negative liberty answers to the question: "What is the area within which the subject — a person or group of persons — is or should be left to do or be what he is able to do or be, without interference by other persons" (Berlin, 1958, pp. 15-16).

Nowadays, within democratic countries, positive liberty has become inversely proportional to negative one. Therefore, it forks: on the one hand, it allows the free exercise of the person through almost infinite self-expression possibilities and technological incentives – "ICTs show people beauties of micro-worlds and of galaxies, and give them immense possibilities to create beauty" (Radovan, 2013,

---

[5] The term purposeful is meant to denote that the act or behavior may be interpreted as directed to the attainment of a goal — i.e., to a final condition in which the behaving object reaches a definite correlation in time or in space with respect to another object or event (Rosenblueth et al., 1943, p. 18).

p. 4) – which spill and multiply in the digital world; on the other hand, it constantly watches over this exercise (e.g., managing IoT, Big Data, and social networks) and this translates in more control and/or information deviation to affect people, such as the case of million Facebook profiles harvested for Cambridge Analytica in major data breach to affect human behavior during political elections (Cadwalladr and Graham-Harrison, 2018). Hence, a negative effect is that omnipresent social web's hyper-connection makes every place essentially like the same and traceable one: the Internet. The hiatus is even more visible from the point of view of the digital divide existing between the Western world and developing countries. Access to information is an important indicator of freedom and, in an increasingly digitized world, having the technical and cultural tools to access it has important repercussions in terms of knowledge. The equation "accessibility to information = knowledge" can therefore be established and it is clear that ad hoc policies can more or less shift the balance of individual liberty. However, this information Era and network economy can generate a global concept of the *Panopticon* theorized by J. Bentham (1748–1832) and then recovered by M. Foucault (1926–1984): it refers to a circular architectural structure, whose centre is occupied by a tower with several large windows, which are opened in front of the internal face of the ring. The peripheral building is divided into different cells filling its entireness. Two windows in it: one inwards, the other one outwards. "Thus, just put a supervisor in the tower and then enclosing a fool, a convict, a workman in every cell and take control over them" (Foucault, 1975, p. 218).

In the globalized world, progress has on the one hand, allowed greater access to (*open*) resources of knowledge that generate the background for greater wellbeing; on the other hand, it has limited individual freedom and increased the gap (and income inequality) between the Western world and developing societies and within these countries[6]. For this reason, it is important to make a distinction between the concepts of "progress" and "evolution" (Gini, 1959, *passim*). The term "progress" underlies an infinite tension – albeit regulatory – towards a perfect society. This approach is ascribable to a theoretical framework of a Platonic mold that conceives of time as linear and is rarely opposed to its opposite, that is, to the concept of "regression". Instead, the term "evolution" is always ascribable to a cycle and therefore it is inserted into a temporal dynamics that implies a return as suggested by the philosopher Vico (Flint, 1884). A linear concept of time, in addition to being difficult to sustain within a complex reality, is blind to the collateral circumstances that occur concurrently with phenomena which by definition are complex.

---

[6] Cf., also Coccia, 2017d for other negative effects of income inequality.

In general, the concept of progress based on economic development with augmentation of wealth and of capital has the fallacy of the identification of the increase of goods with advance toward the socially good. In fact, Barth (1897, p. 296) criticizes Durkheim's theory of the division of labor of Capitalist systems as follows:

> He forgets entirely that moral ideas are ideas about values, and that they cannot hinder progress toward greater wealth of values since they them-selves first fix these values, first create them. A society, for example, permeated by the ascetic morality, might restrict its production; it would nevertheless make no economic retrogression since these diminutions in goods would not be felt as such. Durkheim always assumes that society has no other end than to produce goods.

Economic goods are an important condition of social progress but Woods (1907, p. 810) states that: "human nature, as we know it, is many-sided, and human wants reach out in a multitude of directions toward things which have only a remote relation to economic goods. Any careful definition of progress must take full account of the satisfaction of the social, intellectual, aesthetic, and moral sides of life".

Moreover, how should we interpret the fact that in the most industrialized and HDI countries – and therefore richer, and apparently healthier – there is a higher incidence of cancer than poor countries? Does progress, science and technology driven, always mean happiness and healthy in society? And above all, is happiness quantifiable? From this point of view, the enthusiastic advocates of progress do not take into consideration the side effects of the immoderate and blind economic growth of the new global, post-industrial, and late-capitalistic society: a society driven, more and more, by maximization of profit of large corporations without considering a sustainable development and environment (Coccia, 2012).

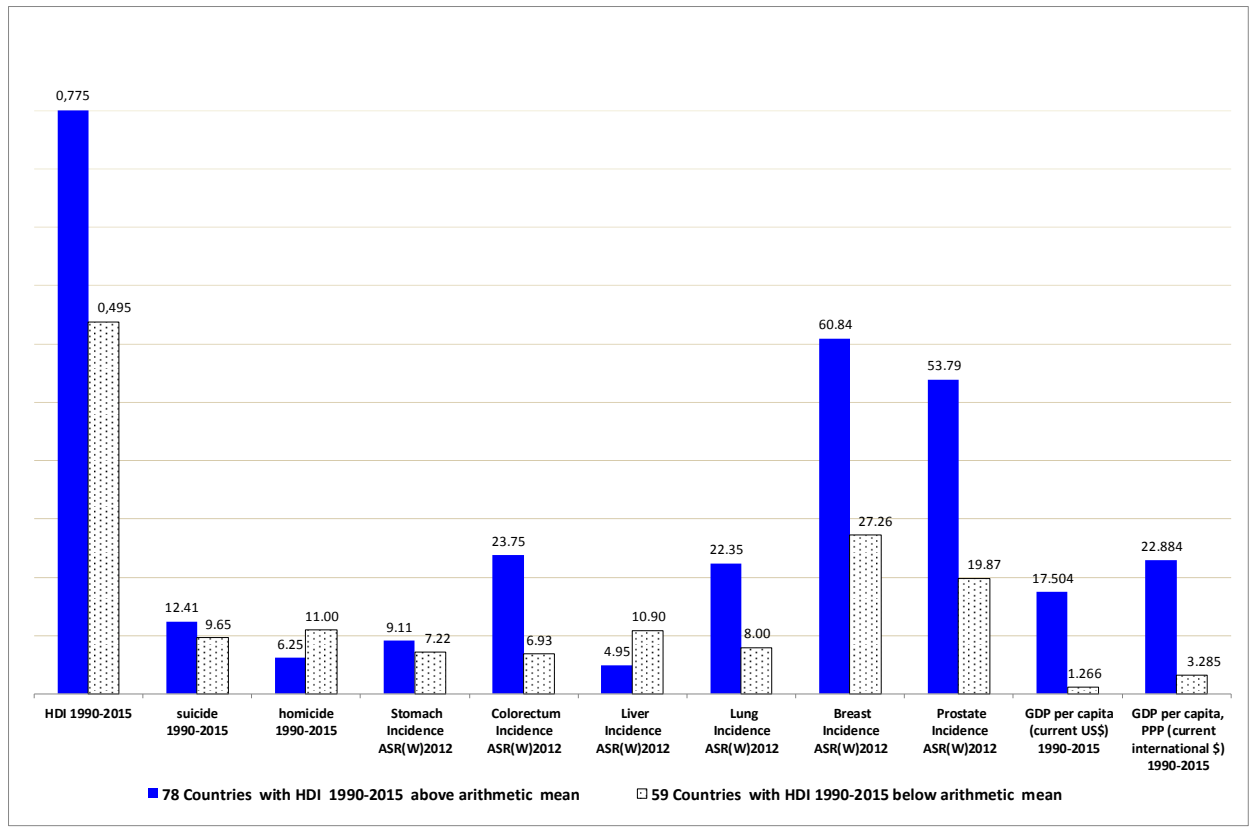

Figure 1. Comparison of indicators of human progress between 137 Countries with Human Development Index (HDI) average 1990-2015 above arithmetic mean (78) and below arithmetic mean (59)

Figure 1 shows an empirical evidence of some negative effects of human progress in society: countries with higher Human Development Index (HDI) have the incidence of some cancer higher than countries with lower HDI. The phenomenon can be explained by examining the previous arguments, i.e. the growth of industrial and technological level of countries, it also increases the risk of pollution with negative repercussions on health. Having said that, and in addition to environmental issues, the highest purchasing power of countries generates benefit from new goods, but in some cases, they may be harmful for health. In fact, Coccia (2015b) shows that technological innovations support human development, which by social mechanisms of population growth, mass production and consumption can also engender pollution and diffusion of some mutagens and genotoxic carcinogens in environment and food chain. For instance, European and US industrialization has generated a general socioeconomic progress and wellbeing in society but also diffusion of pollutants, pesticide in agriculture, several chemicals, asbestos, food processed or chemically preserved, etc. (Coccia, 2015b, p. 62). Put otherwise, progressive development, associated with new needs, has induced a mass production of numerous goods, thereby more consumption in society has damaging effects on environment and people by resources depletion, pollution and diffusion of genotoxic carcinogens (Coccia, 2015b, p. 62). This study by Coccia (2015b) seems to show a main interrelationship between high technological and economic performance[7] (indicators of human progress) and high diffusion of some cancers between countries, controlling screening technology (e.g., computed tomography). In short, results reveal a negative association between diffusion of technology, economic growth (an indicator of progress) and incidence of some cancers in human society (Coccia, 2015b, 2013a, 2012c; Chagpar and Coccia, 2012).

Figure 1 also shows that suicides are higher within countries with higher HDI and this is another evidence to support the thesis that wellbeing and happiness do not derive exclusively from technological and economic progress that lead to better standard of living but rather from its quality not associated to utilitarian (monetary) aspects. Another social issue in figure 1 is the number of murders, which is higher in countries with lower HDI. Homicide, however, is due to socioeconomic factors, such as poverty, income inequality, rather than individual psychosocial risk factors.

---

[7] Cf., Coccia, 2004, 2005a, 2005b, 2009, 2010c.

Table 1. Inconsistencies of human progress: comparison between countries with high and low Human Development Index

| Description variables | Countries with HIGH Human Development Index | Countries with LOW Human Development Index |
|---|---|---|
| Countries | 30 | 30 |
| Human Development Index (HDI), average 1990 - 2015 | 85.64 | 48.56 |
| Suicide, average 2000-2015 | 13.07 | 8.45 |
| Intentional Homicide, average 2000-2015 | 1.45 | 8.66 |
| GINI index of income inequality, average 1990-2005 | 27.81 | 37.05 |
| Energy use (kg of oil equivalent per capita), average 1990-2014 | 52.47 | 4.25 |
| Electric power consumption (kWh per capita), average 1990-2014 | 96.56 | 3.88 |
| Fossil fuel energy consumption (% of total), average 1990-2014 | 78.12 | 37.40 |
| GDP per capita (current US$), average 1990-2015 | 33.69 | 0.97 |
| GDP per capita, PPP (current international $), average 1990-2015 | 35.76 | 2.53 |
| *Incidence cancer ASR (W)* [a] *1-27* | | |
| 1-Lip, oral cavity | 4.17 | 3.42 |
| 2- Nasopharynx | 0.70 | 1.07 |
| 3- Other pharynx | 2.15 | 1.54 |
| 4- Oesophagus | 3.18 | 4.64 |
| 5- Stomach | 8.05 | 6.89 |
| 6- Colorectum | 31.72 | 5.86 |
| 7- Liver | 5.30 | 8.08 |
| 8- Gallbladder | 2.03 | 1.09 |
| 9- Pancreas | 6.55 | 1.68 |
| 10- Larynx | 2.07 | 1.63 |
| 11- Lung | 28.24 | 6.42 |
| 12- Melanoma of skin | 12.25 | 0.84 |
| 13- Kaposi sarcoma | 0.20 | 4.54 |
| 14- Breast | 79.00 | 26.10 |
| 15- Cervix uteri | 7.75 | 27.65 |
| 16- Corpus uteri | 13.08 | 4.35 |
| 17- Ovary | 8.34 | 3.92 |
| 18- Prostate | 77.06 | 17.56 |
| 19- Testis | 6.43 | 0.50 |
| 20- Kidney | 8.05 | 1.02 |
| 21- Bladder | 9.63 | 2.28 |
| 22- Brain, nervous system | 5.22 | 1.79 |
| 23- Thyroid | 7.76 | 1.32 |
| 24- Hodgkin lymphoma | 1.99 | 0.69 |
| 25- Non-Hodgkin lymphoma | 9.21 | 3.70 |
| 26- Multiple myeloma | 2.98 | 0.71 |
| 27- Leukaemia | 7.19 | 2.84 |

*Note*: [a] Incidence data for all ages. Age-standardized rate (W): A rate is the number of new cases per 100,000 persons per year. An age-standardized rate is the rate that a population would have if it had a standard age structure. Standardization is necessary when comparing several populations that differ with respect to age because age has a powerful influence on the risk of cancer. Sources Incidence cancer ASR (W) is: GLOBOCAN (2012) IARC-Section of Cancer Surveillance, http://globocan.iarc.fr/Pages/fact_sheets_population.aspx (24 April 2018).

Table 1 shows detailed data about the inconsistency of human progress, measured by HDI, which generates economic growth (higher GDP per capita and energy consumption) but also many negative sides, such as higher incidence of suicide and cancers.

Hence, the empirical evidence just mentioned supports the criticisms to the equation: human progress = economic growth, wellbeing and happiness. Human progress should be, more and more, directed towards a new model of sustainable environment, food security, sociability and rightness satisfactions of people: i.e., a more attentive and conscious society that critically seizes the present and questions its own future paths.

**Conclusions**

Progress has always been a hot topic of discussion in the Western-style world (Woods, 1907). It is a complex and stratified concept that changes form and specific weight within society depending on historical period and spatial area. We have advanced a definition of human progress that could summarize the different historical trends and generalize the concept over time and space: an ideal of maximum wellbeing driven by achieving new, consequential and progressive goals, endlessly. This definition seems to satisfy four desiderata (Brandon, 1978, p. 189ff): (a) independence; (b) generality; (c) epistemological applicability; and (d) empirical correctness. However, the definition of the concept of human progress in a changing society and so rooted in the present is always a difficult task; our research has the merit of having highlighted and problematized an elusive idea too often given for certain.

The *excursus* on the philosophical debate between Nineteenth and Twentieth centuries allowed us to go to the heart of the problem. These centuries have seen proliferation of theories on progress as never before in history and it is no coincidence that this happened in conjunction with major socioeconomic events and a new economic view of progress (Woods, 1907; Seligman, 1902). Starting from this, the idea that the concept of progress was a cause but also a consequence of the economic vector was strengthened in us. This is why we have emphasized that - at least during the phase of its theoretical systematization - the concept was an ex-post justification. In fact, the Lassalle-Marxian view of progress shows that techno-economic processes and progressive social change are based on readjustment of human institutions and activities to a changing economic environment; as a consequence, human progress consists in the adaptation of life to new economic and social bases (cf., Woods, 1907, pp. 810-811; Bernstein, 1893). Sombart (1898, p. 156-157) claims that:

> History teaches us that what we call advance has always been only change to a higher system of economy, and that those classes thrive who represent this higher system. Behind capitalism there is no "development;" possibly there may be ahead. The degree of production which has been reached by it must in any case be rivalled by any party that will secure the future for itself. In that is shown, I think, the standard of any advance movement

The interwoven relation between economic development and human life is associated with technology that yields a greater satisfaction of human wants at smaller cost than previous technology. This technological substitution generates technological, economic and social change (adjustments). Hence, the underlying factors of economic and social change and as a consequence human progress are human wants and human control of nature by science advances and new technology (Cf. Woods, 1907). De Greef (1895) states that a complete inventory of social activity is necessary for an adequate exhibit of social progress. "Progress in an individual or in a community is thus a function of all the various qualities and aspects of life which are there realized. Not physical well-being alone, nor the abundance of wealth, nor even the moral advance which has been attained, may serve as the measure of progress; all of the interests are required because all are phases of normal human life." (Woods, 1907, p. 817). Moreover, a telic view of progress argues an infinite series of reaccommodations between human experience and human ideals direct to realize fullness of life (Woods, 1907, p. 818). The ultimate form of the criterion of progress must be in terms of the realization of the life of individuals that constitutes the "ultimate social fact" (Woods, 1907, p. 820). Science and technology should be the forerunners of a full realization of the meaning and possibilities of life of individuals. This realization of the life of individual is achieved in appropriate social structures with democratization, good governance, education, culture and sustainable environment. At aggregate level, this goal supports human progress in society. However, these factors of human progress are not always associated to a comprehensive wellbeing and happiness of people as showed in Figure 1 in HDI countries.

Overall, it would be naive to limit the human progress or at least to make it dependent on the economic vector alone. To reiterate, we emphasize that the equation " more progress = more economic growth, wellbeing and happiness" has inconsistencies and is not valid at all because of complex socio-psychological factors affecting human behavior represented by wellbeing, health, curiosity, power, sociability and rightness satisfactions in persons associating (Small, 1905, p. 682).

We conclude that the concept of progress is due to the expanding content of the human life-interests whose increasing realization constitutes progress, rather than external processes conceived in terms of divine will, biological causation, or economic processes, and so on. Human progress is driven by long-run ideal of the essential human interests and endless curiosity that change in society and their satisfaction that characterizes the human nature from millennia (Woods, 1907, p.p. 813-815).

Overall, then, the whole process of human progress is driven by the increasingly effective struggle of the human mind in its efforts to raise superior to the exigencies of the external world and attitude to satisfy human desires, solve problems and achieve/sustain power in a sustainable society. However, a comprehensive definition of human progress, at the intersection of vital elements of economics, sociology, psychology, anthropology, and perhaps biology, is a non-trivial exercise. Even though we could not face a comprehensive analysis of overall characteristics of human nature in our work for its complexity, we believe that psychosocial factors of people in society have their vital weight in the debate on human progress. We assume that an advanced society must support mainly happiness, social wellbeing and sustainable environment, rather than a blind economic growth with consequential environmental, social and food security threats. Future research will explore this *terra incognita* to refine and extend, as far as possible, the concept of human progress in society that possibly is evolved ongoing.